\documentclass[aps,showpacs,prb,twocolumn,superscriptaddress,apsrev]{revtex4-1}

\usepackage{indentfirst}
\usepackage{multirow}
\usepackage{color}
\usepackage{amsfonts}
\usepackage{amsmath}
\usepackage{mathrsfs}
\usepackage{amssymb}
\usepackage{wasysym}
\usepackage{xfrac}
\usepackage{graphicx}
\usepackage{appendix}
\usepackage{CJK}

\usepackage{url}
\usepackage{multirow}

\usepackage{ulem}

\begin{document}

\begin{CJK*}{GBK}{}
\title{Quantum phase diagram for two species hardcore bosons in one-dimensional optical lattices with the resonantly driven Rabi frequency}
\author{Peigeng Zhong}
\affiliation{Beijing Computational Science Research Center, Beijing 100084, China}

\author{Tao Wang}
\email[]{tauwaang@cqu.edu.cn}
\affiliation{Department of Physics, Chongqing University, Chongqing, 401331, China}
\affiliation{Chongqing Key Laboratory for Strongly Coupled Physics, Chongqing University, Chongqing 401331, China}

\author{Shijie Hu}
\email[]{shijiehu@csrc.ac.cn}
\affiliation{Beijing Computational Science Research Center, Beijing 100084, China}

\author{Haiqing Lin}
\affiliation{Beijing Computational Science Research Center, Beijing 100084, China}
\affiliation{Department of Physics, Beijing Normal University, Beijing, 100875, China}

\date{\today}

\begin{abstract}
We propose an experimental realization of the time-periodically modulated Rabi frequency and suggest density-dependent hoppings of two species hardcore bosons in a one-dimensional optical lattice.
Distinct from the previous work [Phys. Rev. Research {\bf 2}, 013275 (2020)], we study effects in the first resonance region.
In the effective Hamiltonian, the intra-species hopping occurs only if the density discrepancy of the other species on these sites is zero, while the inter-species one is allowed once the relevant density discrepancy becomes nonzero.
At integer-$1$ filling, the quantum phase diagram of the effective Hamiltonian is determined by the perturbation analysis together with numerical calculations.
We find that in the limit of dominant $J_{1}$, the system becomes a double-degenerate dimerized state, with spontaneously breaking the translation symmetry.
The interplay of $J_{0}$, $J_{1}$ and the fixed ${\bar U}=1$ leads to three BKT transition lines and a tricritical BKT point.
Exact transition lines are obtained by the level spectroscopic technique.
Besides, general physical properties, including the charge gap, neutral gap, superfluid density and dimerization strength, are investigated as well.
\end{abstract}


\maketitle
\end{CJK*}

\section{Introduction}
Floquet engineering with controllable time-periodic driving sources~\cite{Eckardt_2017,Simonet_2021} is a powerful tool to tailor quantum gases in optical lattices.
In contrast to the traditional solid state systems, it relies on extremely high-intensity laser and a sophisticated scheme of fast measurements~\cite{Cavalleri_2020}.
The typical driving frequency ranges from several to several tens of kilohertz (kHz), which allows us to unveil the physics in the strong driving regime experimentally~\cite{Arimondo_2012}.
Floquet engineering greatly enriches accessible models and exotic phenomena in optical lattices, such as generating artificial gauge fields by breaking the time-reversal symmetry of the driving function~\cite{Windpassinger_2012, Spielman_2012, Schweizer_2019, Barbiero_2019, Lu_2021}, realizing anyon physics combing with Raman assisted tunneling~\cite{Greschner_2015, Eckardt_2016} and designing other unconventional Hubbard models~\cite{Greschner_2014, Zhao_2019}. 

For example, Bose-Hubbard models with density-dependent correlated hoppings have been Floquet-engineered in optical lattices by tuning the magnetic field near a Feshbach resonance point~\cite{Meinert_2016, Compagno_2020}.
These models give novel quantum phases~\cite{Rapp_2012, Wang_2014} and are also essential for implementing dynamical gauge fields~\cite{Reznik_2015, Banuls_2020}.
However, for two species hardcore bosons that tightly connect to the Fermi-Hubbard model in one dimension (1D), it was challenging to tune inter-species interaction together with holding the {\it hardcore} constraint.
In the previous work [Phys. Rev. Research {\bf 2}, 013275 (2020)], we proposed an alternative way of periodically driving the Rabi frequency, which has been realized in experiments recently~\cite{Lu_2021}.
In the case of high-frequency approximation, we found an integrable point and a ``gauge dressed superfluid'' phase~\cite{wang_2020}.
Once the driving frequency resonates with the interaction strength, new physics insights are concerned in the presence of different models~\cite{Simonet_2021,Jaksch_2017,Esslinger_2018}.

The paper is organized as follows.
In section~\ref{sec_HighFExp}, we derive the effective model in the first resonance region according to an existing time-dependent Hamiltonian in optical lattices.
In section~\ref{GroundPD}, we do an explicit perturbation analysis and numerical simulations to characterize all phases and phase transitions in the allowable parameter region.
At last in section~\ref{Summary}, we summarize all results with a brief discussion.

\begin{figure}[t]
\includegraphics[width=0.95\columnwidth]{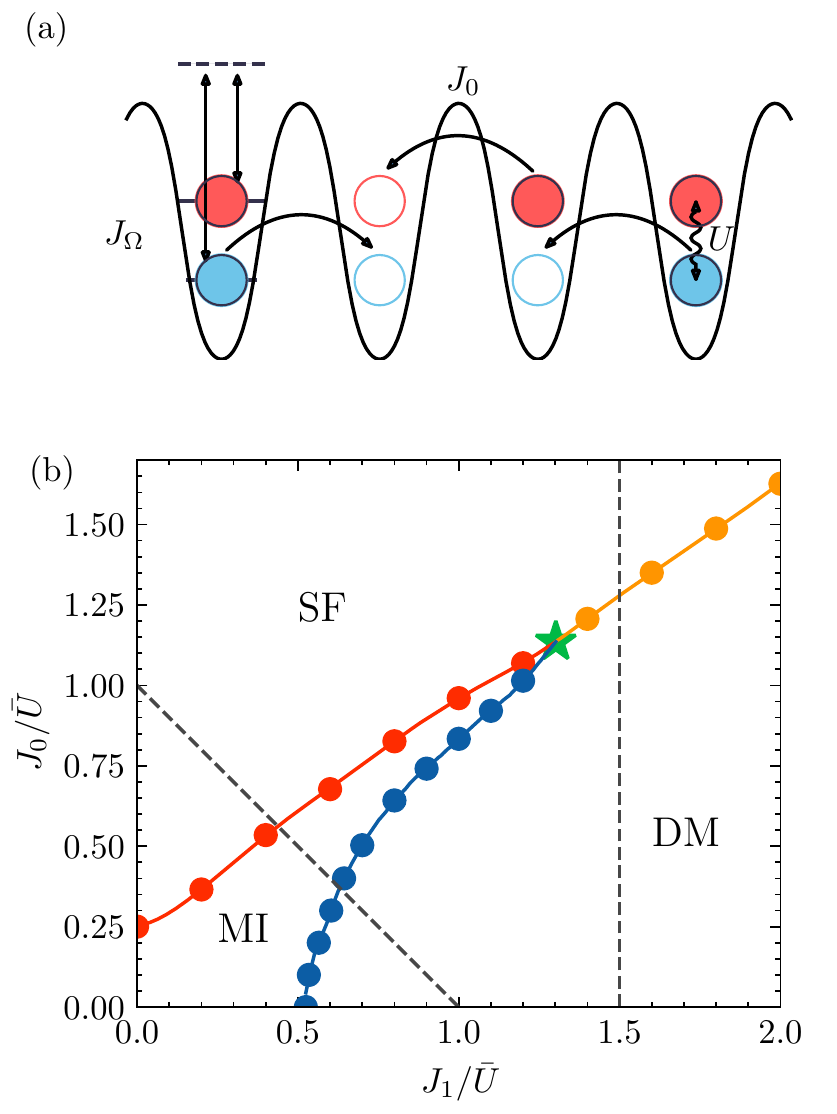}
\caption{(a) Schematic picture for the experimental realization.
Both hopping amplitudes $J_{0}$ and raw onsite repulsion strength are controlled by the lattice depth.
Two hyperfine states are coupled to an intermediate state (black dashed line) by the extra Raman laser beams.
The effective Rabi frequency $J_{\Omega}$ is time-periodically modulated.
The static magnetic field is a bit far away from a Feshbach resonance point on the side of negative scattering length, so the residual inter-species repulsion strength $U\sim\hbar\omega$.
Meanwhile, the dominantly-large intra-species repulsion maintains the hardcore constraint.
(b) Quantum phase diagram of the effective model at integer-$1$ filling. It consists of three distinct phases: superfluid (SF), Mott-insulating (MI) and Dimerized (DM) phases. Border-lines SF-MI, SF-DM and MI-DM are marked by red, orange and blue respectively. A green five-star indicates a tricritical point of three phases. Two dashed cutting lines $J_{0}+J_{1} = 1$ and $J_{1} = 1.5$ are marked for convenience of discussion.
}\label{fig1}
\end{figure}

\section{Density-dependent correlated hoppings in the first resonance region}
\label{sec_HighFExp}
Following the existing scheme~\cite{wang_2020,wang_2021}, the full Hamiltonian for two species hardcore bosons in optical lattices reads
\begin{eqnarray}
{\hat H} (t)&=&{\hat H}_{T}+{\hat H}_{U}+{\hat H}_{\Omega}(t)
\end{eqnarray}
and
\begin{eqnarray}
& &{\hat H}_{T} = -J_{0} \sum_{l} \left({\hat a}^{\dagger}_{l} {\hat a}_{l+1} + {\hat b}^{\dagger}_{l} {\hat b}_{l+1} + \rm{h.c.}\right)\, ,\nonumber\\
& &{\hat H}_{U} = U \sum_{l} {\hat n}^{a}_{l}{\hat n}^{b}_{l}\, ,\nonumber\\
& &{\hat H}_{\Omega}(t) = J_{\Omega}(t) \sum_{l}\left(\hat{a}_{l}^{\dagger}\hat{b}_{l}+\rm{h.c.}\right) \, ,\nonumber\label{H2}
\end{eqnarray}
where ${\hat a}_{l}$ (${\hat b}_{l}$), ${\hat a}^{\dagger}_{l}$ (${\hat b}^{\dagger}_{l}$) and ${\hat n}^{a}_{l}={\hat a}^{\dagger}_{l} {\hat a}_{l}$ (${\hat n}^{b}_{l}={\hat b}^{\dagger}_{l} {\hat b}_{l}$) are the annihilation, creation and particle number operators for the species ``$a$'' (``$b$'') respectively,
$J_{0}$ the hopping coefficient governed by the lattice depth,
$U$ the effective inter-species onsite repulsion strength,
$J_{\Omega}(t)=J^{0}_{\Omega} \cos(\omega t)$ the time-periodically modulated ``Rabi frequency'' with the driving frequency of $\omega$ and the index $l$ runs over whole $L$ lattice sites.

In the previous work, our attention was centered on the non-resonance region, where the driving frequency must be much larger than $J_{0}$ and $U$, but not too large to avoid ``photon-assisted hopping'' between energy bands of the optical lattices~\cite{wang_2020}.
In this work, we further investigate rich quantum phases and phase transitions in the first resonance region $\hbar \omega \sim U$, which is achieved by moving the static magnetic field a bit further away from the Feshbach resonance point on the side of negative scattering length.
Meanwhile, $J^{0}_{\Omega}/\hbar$ is of several kHz, the same as $\omega$.

We are willing to restate three concerns.
Firstly, by using an acousto-optic modulator (AOM) to change both the amplitude and polarization angle of the pump laser~\cite{Eklund_1975, Sapriel_1979}, the setup can effectively generate the time-periodically modulated Rabi frequency in practice~\cite{wang_2020}.
Secondly, the initial phase of the time-periodically modulated Rabi frequency is irrelevant to 
the stroboscopic measurement at an integer number of periods, which has already been realized and well controlled in experiments~\cite{Eckardt_2017}.
Thirdly, although the system usually heats up to the infinite temperature in a long time due to the absence of energy conservation~\cite{Rigol_2014},
a prethermal stage still survives~\cite{Polkovnikov_2016, Rubio-Abadal_2020}.
Various theoretical and experimental studies have exhibited that it is long enough for stabilizing and measuring the effects governed by the effective Hamiltonian~\cite{Porto_2019,Aidelsburger_2020}.
Even in the resonance region~\cite{Esslinger_2018},
the heating rate is exponentially suppressed in the Floquet Bosonic optical lattices~\cite{Messer_2018}.

According to the Floquet theory~\cite{Rapp_2012, Arimondo_2012, Wang_2014, Greschner_2014, Meinert_2016, Eckardt_2017} for a time-periodically driven system ${\hat H}(t)={\hat H}(t+T)$ with a period of $T=2 \pi / \omega$, the dynamics of the system is depicted by the effective Hamiltonian and corresponding kick operators, which can be determined from a nearly-degenerate perturbation method~\cite{Eckardt_2017,wang_2020}.
By adopting standard operations (Appendix~\ref{AppA}), we obtain the effective Hamiltonian
\begin{eqnarray}
{\hat H}_{e} &=& - \sum_{l} \left[J_{0}({\hat H}^{aa}_{l,l+1} + {\hat H}^{bb}_{l,l+1}) + J_{1} ({\hat H}^{ab}_{l,l+1} + {\hat H}^{ba}_{l,l+1})\right]\nonumber\\
& & + {\bar U} \sum_{l} {\hat n}^{a}_{l}{\hat n}^{b}_{l}\label{effham}
\end{eqnarray}
with $J_{1} = J_{0} {\cal J}_{1}(2K_{\Omega})$, the first order of the first kind Bessel function ${\cal J}_{1}(x)$,
a dimensionless parameter $K_{\Omega} = J^{0}_{\Omega} / \hbar\omega$,
a renormlized onsite repulsion strength ${\bar U} = U - \hbar\omega$ and four distinct hopping terms
\begin{eqnarray}
{\hat H}^{aa}_{l,l+1} &\!=\!& {\hat a}^{\dagger}_{l} {\hat a}_{l+1} [{\hat n}^{b}_{l} {\hat n}^{b}_{l+1} + (1-{\hat n}^{b}_{l}) (1-{\hat n}^{b}_{l+1})] \!+\! {\rm h.c.}\, ,\nonumber\\
{\hat H}^{bb}_{l,l+1} &\!=\!& {\hat b}^{\dagger}_{l} {\hat b}_{l+1} [{\hat n}^{a}_{l} {\hat n}^{a}_{l+1} + (1-{\hat n}^{a}_{l}) (1-{\hat n}^{a}_{l+1})] \!+\! {\rm h.c.}\, ,\nonumber\\
{\hat H}^{ab}_{l,l+1} &\!=\!& {\hat a}^{\dagger}_{l} {\hat b}_{l+1} [{\hat n}^{b}_{l} (1-{\hat n}^{a}_{l+1}) + (1-{\hat n}^{b}_{l}) {\hat n}^{a}_{l+1}] \!+\! {\rm h.c.}\, ,\nonumber\\
{\hat H}^{ba}_{l,l+1} &\!=\!& {\hat b}^{\dagger}_{l} {\hat a}_{l+1} [{\hat n}^{a}_{l} (1-{\hat n}^{b}_{l+1}) + (1-{\hat n}^{a}_{l}) {\hat n}^{b}_{l+1}] \!+\! {\rm h.c.}\, .
\end{eqnarray}
Due to the properties of ${\cal J}_{1}$, the intra-species hopping in the effective Hamiltonian occurs only if the density discrepancy of the other species on these sites is zero, while the inter-species one is allowed once the relevant density discrepancy becomes nonzero.
Hereafter we choose ${\bar U}=1$ as unit since two out of three parameters are free in the model.
Additionally, we only consider the case of integer-$1$ filling, namely $N_{a} + N_{b} = L$ and $N_{a/b} = \langle \sum_{l} {\hat n}^{a/b}_{l} \rangle$, where more interesting phenomena and a rich phase diagram are anticipated.

Although $|J_1 / J_0| < 1$ because of $|{\cal J}_{1}| < 1$, its regime could be extended if the driving forms are chosen appropriately.
For instance, one can implement the above proposal in a two-leg Wannier-Stark ladder, where the energy difference between the nearest-neighboring sites along the ladder resonates with the driving frequency $\omega$~\cite{Eckardt_2007}.
In the setup, $J_0$ and $J_{1}$ are dressed by ${\cal J}_{1}$ and the second order of the first kind Bessel function ${\cal J}_{2}$ separately, so we study the full parameter space of $|J_{0} /J_{1}|$ in the following sections.

Before going to the details of the quantum phase diagram, let us discuss the symmetry related to the controllable parameters $J_{0}$, $J_{1}$ and ${\bar U}$.
First of all, positive and negative $J_{0/1}$ are linkable under the transformation ${\cal W}^{a/b}_{l} = \exp\left(i\pi{\hat n}^{a/b}_{l}\right)$, which yields ${\hat a}_{l} \rightarrow -{\hat a}_{l}$ and ${\hat b}_{l} \rightarrow -{\hat b}_{l}$ respectively.
For instance, in order to reverse the sign of $J_{0/1}$ at the same time, we can apply ${\cal W}^{a/b}_{l}$ to all even sites.
Secondly, the effective Hamiltonian does not have the particle-hole symmetry for each species in principle.
Under the partial particle-hole transformation with ${\hat a}^{\dagger}_{l} \rightarrow {\hat a}_{l}$ for each site, the hopping terms remain invariant but ${\hat H}_{\bar U}$ becomes ${\hat H}_{-{\bar U}} + {\bar U} \sum_{l} {\hat n}^{b}_{l}$.
In the quantum phase diagram discussed later, phases always have zero polarization $N_{a} = N_{b}$ and the chemical potential term ${\bar U} \sum_{l} {\hat n}^{b}_{l}$ is a trivial constant actually.
Thus, the positive and negative $\bar U$ regions are symmetrically equivalent.
In the work, we are merely concerned with the region where both $J_{0/1}$ and ${\bar U}$ are positive.

\section{Quantum phase diagram}
\label{GroundPD}

\subsection{Perturbation analysis}
Once $J_{1} = 0$, the hopping between neighboring double occupied and empty sites is not allowed~\cite{wang_2020}.
The Hamiltonian no longer distinguishes between double occupied and empty sites, we can denote both of them with pseudo-spin up $|\!\!\uparrow\rangle$.
Similarly, the hopping between neighboring single occupied sites is forbidden regardless of whether they are species ``$a$'' or ``$b$'', so both can be denoted with pseudo-spin down $|\!\!\downarrow\rangle$.
The resulting Hamiltonian at integer-$1$ filling is equivalent to an exactly integrable spin-$1/2$ XY model
\begin{eqnarray}
{\hat H}_{J_0} = \sum_{l} \left[ -J_{0} ({\hat S}^{+}_{l} {\hat S}^{+}_{l+1} + {\rm h.c.}) + \frac{\bar U}{2} ({\hat S}^{z}_{l} + 1 / 2) \right]\ ,
  \label{eq:1st}
\end{eqnarray}
where ${\hat S}^{\pm/z}$ are the ordinary operators for spin-$1/2$.
$\bar U$ provides a Zeeman Splitting between $|\!\!\uparrow\rangle$ and $|\!\!\downarrow\rangle$.
For ${\bar U}>4J_{0}$, the system is saturated with only single occupied sites and has a finite charge gap, which corresponds to the Mott-insulating (MI) phase.
When ${\bar U}\leq 4J_{0}$, the groundstate is in a gapless $xy$ phase {\it without\/} superfluid (SF) response indicated by a vertical line $J_{1}=0$ in Fig.~\ref{fig1}~(b).

We now turn to the case of $J_{0}=0$.
When $J_{1} = 0$ as well, the system falls into a clean MI state purely made up of single-occupied sites.
Given a small $J_{1}$, cooperative hoppings are allowed, where the annihilation of a double occupancy at site-$l$ is followed by creating one at site-$(l+1)$, namely $|a\rangle_{l} |a\rangle_{l+1} \rightarrow |0\rangle_{l}|ab\rangle_{l+1}$ and vice verse.
By the second order perturbation, the above hoppings are generalized into an effective Hamiltonian (Appendix~\ref{AppB})
\begin{eqnarray}
{\hat H}^{(2)}_{\text{MI}} \!=\! -\frac{J^{2}_{1}}{\bar U} \sum_{l} \left[ 2({\hat S}^{+}_{l} {\hat S}^{+}_{l+1} \!+\! {\rm h.c.}) \!-\! 4 {\hat S}^{z}_{l} {\hat S}^{z}_{l+1} + 1 \right]\ .
  \label{eq:2rd}
\end{eqnarray}
Different from the Hamiltonian (\ref{eq:1st}), the spin states $|\!\!\uparrow\rangle$ and $|\!\!\downarrow\rangle$ here correspond to the single occupation $|a\rangle$ and $|b\rangle$ respectively.
We further write a new set of bases following $|\!\!\uparrow\rangle \rightarrow - |\!\!\downarrow\rangle$ and $|\!\!\downarrow\rangle \rightarrow - |\!\!\uparrow\rangle$ at all odd sites. Accordingly, the effective hamiltonian goes back to a regular form of the antiferromagnetic Heisenberg chain
\begin{eqnarray}
{\hat H}^{(2)}_{\text{MI}} = \frac{J^{2}_{1}}{\bar U} \sum_{l} \left( 4{\hat {\mathbf S}}_{l} \cdot {\hat {\mathbf S}}_{l+1}  - 1 \right)\ .\label{ohei}
\end{eqnarray}
In the light of the Bethe Ansatz solution~\cite{Lieb_1968}, we know the energy per site $e^{(2)}_{MI} = -4 (\ln2) (J^{2}_{1}/{\bar U})$ in the thermodynamical limit (TDL).

Usually, a regular chain of two species interacting hardcore bosons favors establishing a counter-flow SF order in the MI phase~\cite{Kuklov_2003, Altman_2003, Kuno_2013, Venegas-Gomez_2020}.
However, the $J_{1}$ density-dependent hoppings in the Hamiltonian (\ref{effham}) give a non-local quadrupole order~\cite{Zvyagin_2019, Gen_2019}, which easily breaks into a polarized fluid under the species-dependent perturbation.

When $J_{1} \gg {\bar U}$ and $J_{0}=0$, we suppose that a pure DM state is a product of the local dimers $|\phi_{l,l+1}\rangle = B[ 4 J_{1} (|ab\rangle_{l} |0\rangle_{l+1} + |0\rangle_{l} | ab \rangle_{l+1}) + (\bar{U}+A) (| a \rangle_{l}| a \rangle_{l+1} + | b \rangle_{l}| b \rangle_{l+1}) ]$ on all even bonds.
And the average energy per site is $e^{(0)}_{DM}=({\bar U} - A)/4$ with $A=\sqrt{16J^{2}_{1}+{\bar U}^2}$ and $B = 1/(2\sqrt{A^{2} + {\bar U}A})$.
According to the details in Appendix~\ref{AppB}, we estimate the modification of average energy up to the second order, that is $e^{(1)}_{\text{DM}}=0$ and $e^{(2)}_{\text{DM}}=(J^{2}_{1} / 4) [64 J^{4}_{1} + {\left({\bar U} + A\right)}^4] / {[16J^{2}_{1} + {\bar U}({\bar U} + A)]}^2/({\bar U}-A)$.

The competition between the MI and DM phases leads to a critical point, roughly indicated by $e^{(2)}_{\text{MI}} (J^{c}_{1}) = e^{(0)}_{\text{DM}} (J^{c}_{1}) + e^{(2)}_{\text{DM}} (J^{c}_{1})$, which suggests $J^{c}_{1}\approx 0.3$.
In comparison with $J^{c}_{1} \approx 0.5$ determined by numerical simulations later as shown in Fig.~\ref{fig1}~(b), the underestimated value above-mentioned is caused by the limited perturbation order near the transition point.

Now we do a similar analysis on the charge gap in the MI phase keeping $J_{0/1} \ll {\bar U}$.
And the signal of the gap closure implies the SF-MI transition.
Adding an atom ``$a$'' to the clean MI state immediately raises the repulsion energy.
Meanwhile, the added atom can also move over long distances.
In the perturbation theory (Appendix~\ref{AppB}), the relevant projection operator onto the clean MI state reads
\begin{eqnarray}
{\hat {\cal P}}_{\text{MI}+}=\sum_{l} {\hat {\cal P}}^{(d)}_{l} \sum_{\{r,q\ne l\}} \prod_{r} {\hat {\cal P}}^{(a)}_{p} \prod_{q}{\hat {\cal P}}^{(b)}_{q},
\end{eqnarray}
with operators ${\hat {\cal P}}^{(d)}_{l} = {\hat n}^{a}_{l} {\hat n}^{b}_{l}$, ${\hat {\cal P}}^{(a/b)}_{l} = {\hat n}^{a/b}_{l} (1- {\hat n}^{b/a}_{l})$,
$l$ running over all double occupied sites, $p$ and $q$ giving a complete set of combinations of the single occupied sites ``$a$'' or ``$b$''.
Only up to the first order, the $J_{1}$ terms contribute none at all, while $J_{0}$ allows the exchange between a double occupation and a single one at neighboring sites.
The resulting effective Hamiltonian is
\begin{eqnarray}
& &{\hat H}^{(1)}_{\text{MI}+} = -J_{0} \sum_{l} {\hat {\cal P}}^{(s)}_{l} \left[{\hat H}^{aa,1}_{l,l+1} + {\hat H}^{bb,1}_{l,l+1} \right]\ ,
\end{eqnarray}
with the single-occupation projection operator ${\hat {\cal P}^{(s)}}_{l} = \sum_{r,q\ne l, l+1} \prod_{r} {\hat P}^{(a)}_{r} \prod_{q}{\hat P}^{(b)}_{q}$, ${\hat H}^{aa,1}_{l,l+1} = {\hat a}^{\dagger}_{l} {\hat a}_{l+1} {\hat n}^{b}_{l} {\hat n}^{b}_{l+1} + {\rm h.c.}$ and ${\hat H}^{bb,1}_{l,l+1} = {\hat b}^{\dagger}_{l} {\hat b}_{l+1} {\hat n}^{a}_{l} {\hat n}^{a}_{l+1} + {\rm h.c.}$.
In such a complicated Hamiltonian, we find that the groundstate has highly-degenerate manifolds corresponding to various combinations of the single occupied sites regardless of ``$a$'' or ``$b$''.
For simplicity, we just choose the case where all sites are occupied by ``$b$" except for the site $l$.
So the Hamiltonian becomes
\begin{eqnarray}
{\hat H}^{(1)}_{\text{MI}+} = -J_{0} \sum_{l} \left( {\hat a}^{\dagger}_{l} {\hat a}_{l+1} + {\rm h.c.}\right),
\end{eqnarray}
which yields a groundstate energy $e^{(1)}_{\text{MI}+}=-2J_{0}+{\bar U}$ in the TDL.
Likewise, removing an atom gives the kinetic energy $e^{(1)}_{\text{MI}-}=-2J_{0}$. 
Consequently, we get the charge gap $\Delta_{c} = e^{(1)}_{\text{MI}+} + e^{(1)}_{\text{MI}-}  - 2 e^{(2)}_{\text{MI}} = -4 J_{0} + {\bar U} +8(\ln2)(J^{2}_{1}/{\bar U})$.
The SF-MI borderline $\Delta_c = 0$ or equivalently $J^{c}_{0} = {\bar U}/4 + 2(\ln 2)(J^{2}_{1}/{\bar U}) \propto J^{2}_{1}$ gives the asymptotical behavior near $J_{1}=0$ in Fig.~\ref{fig1}~(b).

At the end of the section, we roughly estimate the borderline from the SF to DM phases for the case of large $J_{0/1}$ but relatively small ${\bar U}$.
As $J_{0} \gg J_{1}$, the system recovers integrability and the groundstate energy is roughly $-2J_{0} /\pi$.
In contrast, following equations for $e^{(0/2)}_{\text{DM}}$ as $J_{0} \ll J_{1}$, the groundstate energy gives $-69J_{1}/64$.
As a result, the SF-DM borderline is $J_{0} / J_{1} = 69 \pi/ 128\approx1.69$.

\subsection{Numerical results}
Beyond the above perturbation analysis, we look forward to an elaborate description of the quantum phase diagram systematically through numerical simulations.

In the phase diagram, the transition from the SF to MI phases breaks neither a continuous nor a discrete symmetry and is clearly of the BKT type in the $1$D quantum system.
Nevertheless, both the MI-DM and SF-DM transitions take place by breaking the translation symmetry, so the criticality depends on the scaling dimensions of relevant operators.
From a trial (Appendix~\ref{AppC}), the dimerization strength does not follow the regular finite-size scaling with a polynomial function and shows us a clear behavior of the imbalanced data collapse instead.
Therefore we believe that both of them are also of the BKT type.

Even though it is very interesting to understand the mechanism by microscopic theories, e.g. bosonization, we only exploit the level spectroscopy technique to determine the BKT transition lines here.
Level spectroscopy technique, based on the renormalization group analysis and the symmetry consideration, was proposed in history to overcome the difficulties of determining the BKT phase transitions~\cite{Nomura_1998, Matsuo_2006, Ueda_2021}.
The BKT transition points have been proven to be related to the crossings of representative excited energy levels following the quantum field theory~\cite{Nomura_1998, Matsuo_2006, Ueda_2021}.

\begin{figure}[t]
\includegraphics[width=0.95\columnwidth]{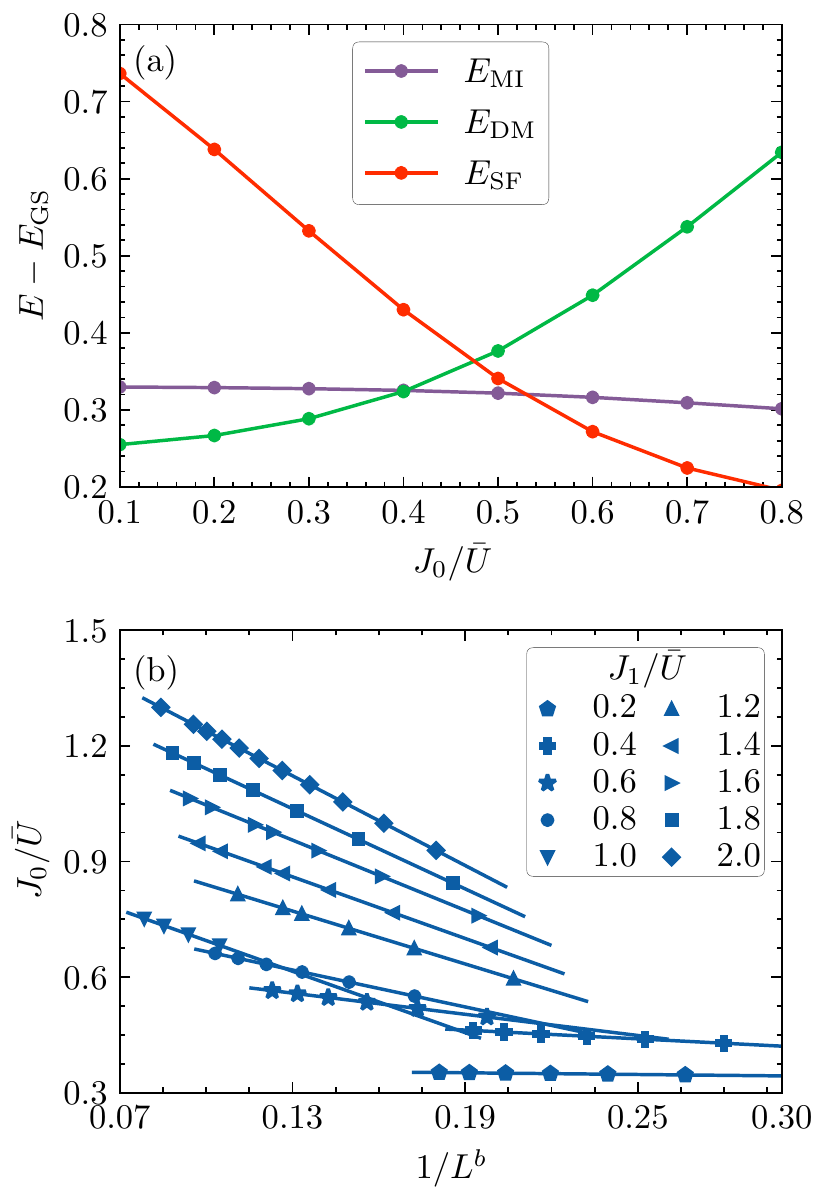}
\caption{BKT transition points determined by the level spectroscopy technique. Each quasi-critical point is marked by the intersection of relevant representative states for a certain $L$: $E_{\text{MI}}$ (violet), $E_{\text{DM}}$ (green) and $E_{\text{SF}}$ (red) for MI, DM and SF respectively. (a) Take an example for fixed $L=14$ and $J_{1}/{\bar U}=0.8$, their intersections give quasi-critical points located at $J^{c}_{0} / {\bar U}=0.403(5)$ for the MI-DM transition and $0.526(5)$ for the SF-MI transition. (b) Scaling of borderlines of SF phase for $J_{1} / {\bar U}$ from $0.2$ to $2.0$, where $L \le 36$ and interval of physical parameters $\delta J_{0/1} /{\bar U}\ge 0.005$. Values of  exponent $b$ by fitting (blue lines) are listed in TABLE \ref{criticalexponent}.}\label{fig2}
\end{figure}

Under the periodic boundary condition (PBC), the $s$-th ($0$, $1$, $\ldots$) lowest-lying level is marked by the number of atoms $N=N_{a}+N_{b}$, momentum $k$ and parity $P$, the energy of which is referred to as $E_{s} (N, k, P)$ in ascending order.
The groundstate $E_{\text{GS}} = E_{0}(L, 0, +1)$ at integer-$1$ filling always has zero momentum and even parity.
In the DM phase, a higher-energy state $E_{\text{DM}} = E_{0}(L, \pi, +1)$ has a small gap for a finite $L$ and gradually becomes degenerate with the groundstate in the TDL.
In the MI phase, the state $E_{\text{MI}} = E_{1} (L, 0, +1)$ describes the gapless excitation with zero neutral gap $\Delta_n=E_{\text{MI}}-E_{\text{GS}}$ in the TDL.
Once $J_{0}=0$, it forms a triplet excitation together with states $E_{2} (L, 0, +1)$ and $E_{0} (L, \pi, -1)$.
In the SF phase, the states with one atom more or less form the charge excitations.
A bit specially, because the ${\bar U}$ term does not have the particle-hole symmetry, we use the average value $E_{\text{SF}} = \min_{k,P}[E_{0} (L + 1, k, P) + E_{0} (L - 1, k, P)]/2$ in practice.

In the first step, we obtain all representative levels by the exact diagonalization (ED) and DMRG methods, followed by getting the quasi-critical points $J^{c}_{0}(L)$ for a finite $L$.
We take the SF-MI and SF-DM transitions as examples.
For $L\le36$, the minimum spacing of parameters $J_{0/1} / {\bar U}$ is equal to $0.005$.
In DMRG, the truncated dimensions $m \le 4096$ and the number of sweeps is enough for convergence accuracy.
In the vicinity of quasi-critical points, we adopt a cubic interpolation of the raw data before accurately determining the places of crossings.

In the second step, we use data extrapolation to estimate the true critical points in the TDL following a polynomial function of $1/L^{b}$.
The values of exponent $b$ in Fig.~\ref{fig2} are listed below.
\begin{table}[ht]
\caption{Values of exponent $b$ fitting in Fig.~\ref{fig2}}\label{criticalexponent}
\centering
\begin{tabular}{c | r | r | r | r | r | r | r | r | r | r |}
\hline\hline
$J_{1} / {\bar U}$ & $0.2$ & $0.4$ & $0.6$ & $0.8$ & $1$ \\ 
\hline
$b$       & $0.476986$ & $0.459362$ & $0.584881$ & $0.633980$ & $0.710094$ \\
\hline\hline\hline
$J_{1} / {\bar U}$ & $1.2$ & $1.4$ & $1.6$ & $1.8$ & $2$ \\
\hline
$b$ & $0.634221$ & $0.649848$ & $0.658709$ & $0.677195$ & $0.690281$\\
\hline\hline
\end{tabular}%
\label{table:exponent}
\end{table}

In the DM phase, each dimer $|\phi_{l,l+1}\rangle$ is almost frozen and can not move easily.
$J_0$ term tends to destroy a local dimer and form a running $ab$ pair in the chain.
Therefore, between the SF and DM phases, it expects to create a paired SF (PSF) with a finite charge gap $\Delta_c$ but a zero neutral gap in principle.
In that case, as $J_0$ increases, $E_{\text{DM}}$ would first intersect with the PSF state $E_{\text{PSF}}=E_{1} (L, 0, +1)$ and then $E_{\text{SF}}$.
In fact, we find that $E_{\text{DM}}$ just has single crossing with $E_{\text{SF}}$.
So there is no evidence of the appearance of the PSF phase at all.
Besides, a tricritical point marked by a five star in Fig.~\ref{fig1}~(b), where all representative states merge, is trivial without any new emerging symmetry.

\begin{figure}[t]
\includegraphics[width=0.95\columnwidth]{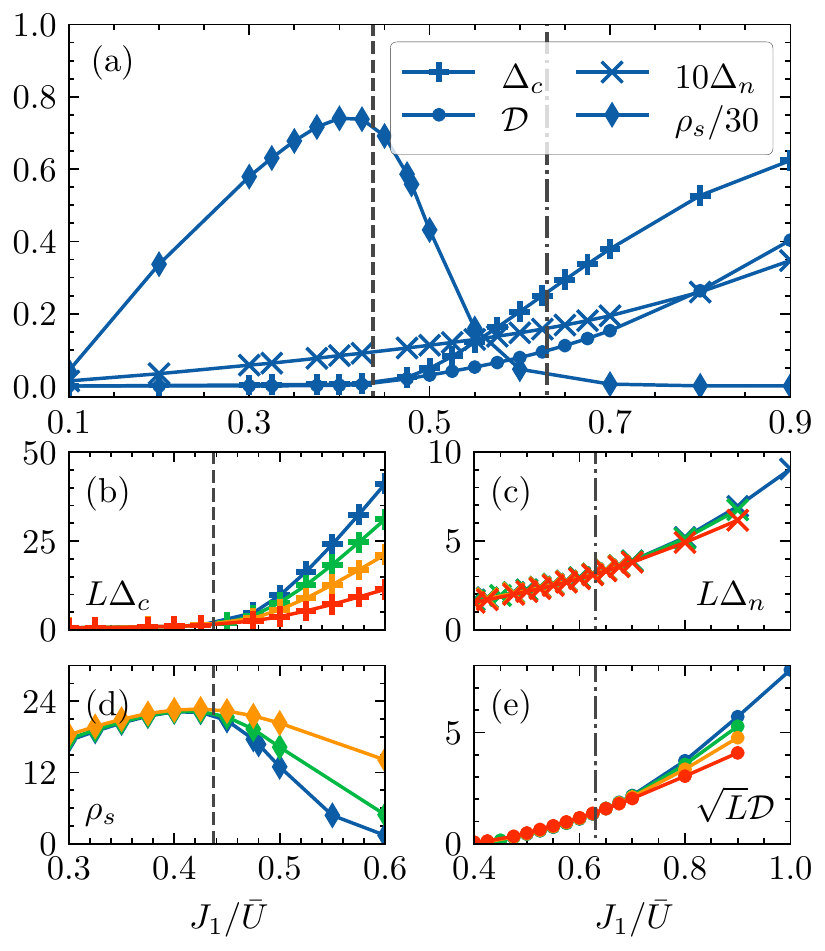}
\caption{(a) Charge gap $\Delta_{c}$ ($+$), neutral gap $\Delta_{n}$ ($\times$), superfluid density $\rho_{s}$ ($\bullet$) and dimerization strength $\cal D$ ($\blacklozenge$) as a function of $J_{1}/{\bar U}$ along a line $J_{1} + J_{2} = {\bar U}$. Left dashed and right dot-dashed lines indicate two BKT transition points $J^{c}_{1}=0.438(4)$ and $0.63(2)$ determined by level spectroscopy technique previously.
Scaled charge gap $L\Delta_{c}$ (b), scaled neutral gap$L\Delta_{n}$ (d), superfluid density $\rho_s$ and scaled dimerization $\sqrt{L} {\cal D}$ (e) for various $L$ are zoomed in near transition points. In details, $\Delta_{c}$, $\Delta_{n}$ and ${\cal D}$ are obtained under OBC for $L=50$ (red), $100$ (orange), $150$ (green) and $200$ (blue) respectively. However, $\rho_s$ is calculated under PBC for $L=10$ (orange), $20$ (green) and $30$ (blue) respectively.
}\label{fig3}
\end{figure}

We now follow two dashed cutting lines in Fig.~\ref{fig1}~(b) to browse the properties of phases.
The dashed line $J_{0} + J_{1} = 1$ or equivalently $J_{0} = 1 - J_{1}$ goes through three distinct phase regions.
For a small $J_{1}$ close to the $J_{0}$ axis, the system stays in the SF phase because of $J_{0} > {\bar U}/4$ and the charge gap closes following a function of $1/L$, which means that $L\Delta_{c} $ collapses to a constant as $J_{1} / {\bar U} < 0.44(5)$ shown in Fig.~\ref{fig3}~(a) and (b) respectively~\cite{Blote_1986, Affleck_1988, Rossini_2012}.
Under a twisting angle $\theta$ at edges, the system has a finite energy response $\delta E_{\text{GS}}(\theta) = E_{\text{GS}}(\theta) - E_{\text{GS}}(0)$.
In the SF phase, the second order differential $\rho_s = \delta E_{\text{GS}}(\theta) / \theta^2$, so called the ``superfluid density''~\cite{Prokofev_2000, Masaki-Kato_2019}, remains finite as $\theta$ decreases towards infinitesimal.
In Fig.~\ref{fig3}~(d), $\rho_s$ is finite and invariant with respect to $L$ in the deep SF region.
While near the SF-MI transition, $\rho_s$ decreases as $L$ grows and expects to vanish in the TDL.
Besides, in the SF phase region, the lowest-lying neutral excitation $\Delta_n = 2 \Delta_c$, considered as the second order effect of the charge excitations correspondingly, also scales of $1/L$ and vanishes in the TDL.

Differently, when the system entries the MI phase region, the charge gap $\Delta_c$ becomes finite and the scaled value $L\Delta_c$ expects to be divergent in TDL.
So in Fig.~\ref{fig3}~(b), the single line starts to split into diverse ones at the SF-MI transition point $J_{1}=0.44(5)$.
However, the neutral gap remains closed, which means that a pair of neighboring $a$ can be flipped to a pair of neighboring $b$ or vice versa without costing energy.
Likewise, the scaled neutral gap $L\Delta_n$ is a constant shown in Fig.~\ref{fig3}~(c).

\begin{figure}[t]
\includegraphics[width=0.95\columnwidth]{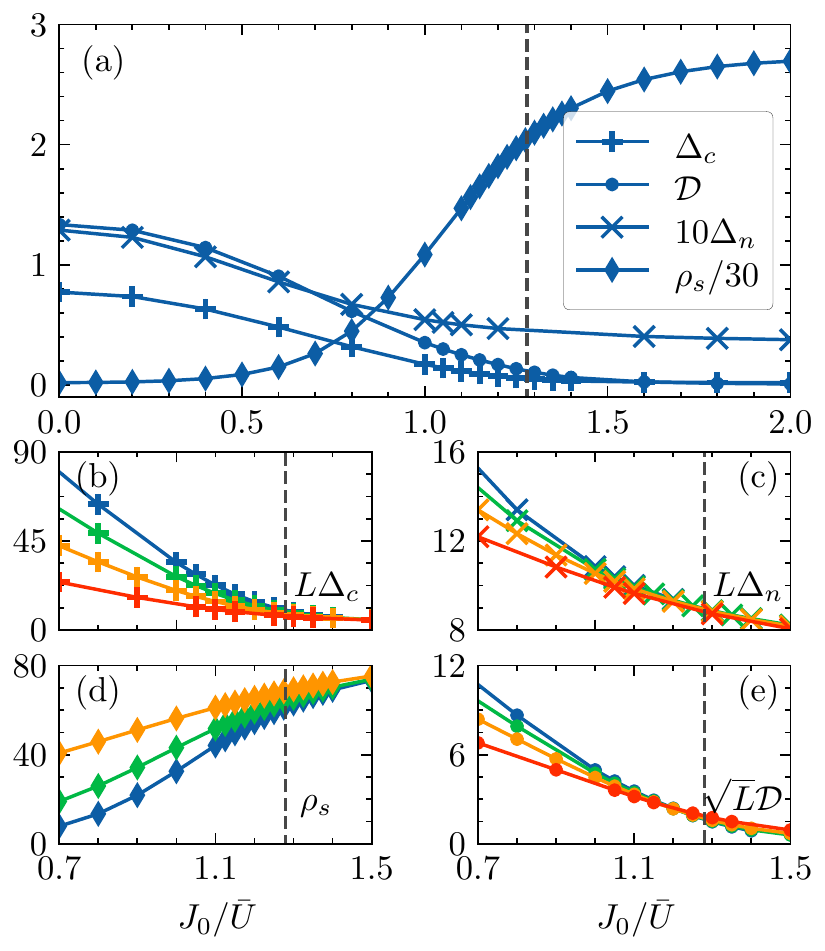}
\caption{
(a) Charge gap $\Delta_{c}$ ($+$), neutral gap $\Delta_{n}$ ($\times$), superfluid density $\rho_{s}$ ($\bullet$) and dimerization strength $\cal D$ ($\blacklozenge$) as a function of $J_{1}/{\bar U}$ along a line $J_{1} / {\bar U} = 1.5$. The dashed line indicates the BKT transition point $J^{c}_{0}=1.28(4)$ determined by level spectroscopy technique previously.
Scaled charge gap $L\Delta_{c}$ (b), scaled neutral gap$L\Delta_{n}$ (d), superfluid density $\rho_s$ and scaled dimerization $\sqrt{L} {\cal D}$ (e) for various $L$ are zoomed in near transition points. In details, $\Delta_{c}$, $\Delta_{n}$ and ${\cal D}$ are obtained under OBC for $L=50$ (red), $100$ (orange), $150$ (green) and $200$ (blue) respectively. However, $\rho_s$ is calculated under PBC for $L=10$ (orange), $20$ (green) and $30$ (blue) respectively.
}\label{fig4}
\end{figure}

In the DM phase, both the charge and neutral gaps are finite.
For an even number of sites with the PBC or an odd number with the open boundary condition (OBC), the groundstate is double-degenerate due to the broken translation symmetry.
As $J_{1}>0.63(5)$ in Fig.~\ref{fig3}~(a), the finite strength discrepancy between even and odd bonds is characterized by the nonzero dimerization strength ${\cal D} = |\langle {\hat h}_{2l,2l+1}\rangle - \langle {\hat h}_{2l-1,2l}\rangle| \ne 0$ with the local Hamiltonian ${\hat h}_{l,l+1} = -J_{0} ({\hat H}^{aa}_{l,l+1} + {\hat H}^{bb}_{l,l+1}) -J_{1} ({\hat H}^{ab}_{l,l+1} + {\hat H}^{ba}_{l,l+1})$ for bond-$l$.
But in the MI and SF regions, $\sqrt{L}{\cal D}$ turns a constant in the TDL shown in Fig.~\ref{fig3} (e).
Referring to the Peierls theory about the DM phase~\cite{Haldane_1982, White_1996, Hu_2014}, changes in the behavior of $\sqrt{L}{\cal D}$ imply the MI-DM transition, the place of which matches the value determined by the level spectroscopy technique previously.

Following the other dashed line $J_1=1.5$, the groundstate transits from the DM phase to the SF phase with the closure of both charge and neutral gaps, as shown in Fig.~\ref{fig4}~(a), which was also verified by the scaling behaviors of $\Delta_c$, $\Delta_n$, $\rho_s$ and $\cal D$ in Fig.~\ref{fig4}~(b)-(e).
Moreover, the strong finite-size effects near $J^{c}_{0} = 1.28(4)$ result in an interesting SF-DM coexistence region $0.6 \lesssim J_{0} < J^{c}_{0}$ for $L=200$, indicated by both finite $\rho_{s}$ and $\cal D$.
It vanishes in the TDL but may exist in other lattice models for the ultra-cold atoms.

\section{Summary and Discussion}
\label{Summary}
Our work provides a way of Floquet engineering a new example in the family of correlated-hopping Bose-Hubbard models.
We propose to load two species hardcore bosons into 1D optical lattices with the time-periodically modulated Rabi frequency.
In the first resonance region, the inter-species and intra-species hoppings of the effective Hamiltonian are adjusted independently by tuning the driving amplitude.
Particularly, the intra-species hopping occurs only if the density discrepancy of the other species on these sites is zero, while the inter-species one is allowed once the relevant density discrepancy becomes nonzero.
In the effective model, $J_{0}$ is only related to the lattice depth, while the ratio of $J_{0}$ to $J_{1}$ can be adjusted by the driving frequency $\omega$.
In addition, the static magnetic field controls the renormalized repulsion strength $\bar U$.

The quantum phase diagram at integer-$1$ filling is studied by the perturbation analysis and numerical simulations.
On condition that both $J_{0}$ and $J_{1}$ are very small, the groundstate favors the MI phase with a finite charge gap and a zero neutral gap.
As $J_{1} = 0$, an integrable model emerges with a highly-degenerate incoherent state.
Furthermore, small $J_{1}$ lifts the degeneracy in the presence of the SF phase with zero charge and neutral gaps as well as finite superfluid density.
A double-degenerate DM state breaks the translation symmetry as $J_{1} / J_{0} = {\cal J}_{1} [2K_{\Omega}] > 0.5$ allowable in the new setup.
In the DM phase, both charge and neutral gaps keep open.
The interplay of $J_{0}$, $J_{1}$ and the fixed ${\bar U}=1$ leads to three BKT transition lines and a tricritical BKT point, the exact places of which can be determined by the level spectroscopic technique.

Many future works could be done based on the effective model.
For example, it is worth studying quantum scars near the integrable point.
Also in a two-leg Wannier-Stark ladder, the frozen dimers in the DM phase would be released and their collective breathing tends to form a resonating valence bond state in a proper parameter space, which certainly deserves more systematic studies in the future.

\section{Acknowledgement}
We thank Xuefeng Zhang, and Xiaoqun Wang for the grateful discussion.
T. W. acknowledges funding from the National Science
Foundation of China under Grants No. 11804034, No. 11647165, Grant No. 12147102, and China Postdoctoral Science Foundation Funded Project No. 2020M673118.
P. G. Z., and S. H. acknowledge funding from the National Science
Foundation of China under Grants No. 12174020.
P. G. Z., S. H., and H. L. further acknowledge support from Grant NSAF-U1930402.
The computations were performed on the Tianhe-2JK at the Beijing Computational Science Research Center (CSRC) as well as Quantum Many-body {\rm I} cluster, at the School of Physics and Astronomy, Shanghai Jiaotong University.

\appendix
\section{Effective Hamiltonian in the first resonance region}\label{AppA}
At the preliminary step of the standard operations, we should remove extra energy scales $J^{0}_{\Omega}$ and $U$ by applying a time-dependent rotation~\cite{Eckardt_2017}, both of which are comparable with $\hbar \omega$.
The rotation ${\hat V} (t) = {\hat V}_{\omega} (t) {\hat V}_{\Omega} (t)$ consists of two commuting parts
\begin{eqnarray}
{\hat V}_{\omega} (t) &=& \exp \left[ -i \omega t \sum_{l} {\hat n}^{a}_{l} {\hat n}^{b}_{l} \right]\, ,\nonumber\\
{\hat V}_{\Omega} (t) &=& \exp \left[ -i {\tilde K}_{\Omega} (t) \sum_{l}\left(\hat{a}_{l}^{\dagger}\hat{b}_{l}+\rm{h.c.}\right) \right]
\end{eqnarray}
with dimensionless parameters ${\tilde K}_{\Omega}(t)=K_{\Omega}\sin(\omega t)$ and $K_{\Omega} = J^{0}_{\Omega} / \hbar\omega$.
After rotation, the Hamiltonian is divided into two parts ${\hat H}^{r}(t) = {\hat H}^{r}_T(t) + {\hat H}_{\bar U}$, where
\begin{eqnarray}
\hat{H}^{r}_{T}(t) &=& - J_{0} \sum_{l} \left[ {\hat J}^{b}_{l,l+1} {\hat a}^{\dagger} _{l} {\hat a}_{l+1} + {\hat J}^{a}_{l,l+1} {\hat b}^{\dagger} _{l} {\hat b}_{l+1} \right. \nonumber\\
& &\left. + {\hat Q}^{ba}_{l,l+1} {\hat a}^{\dagger}_{l} {\hat b}_{l+1} + {\hat Q}^{ab}_{l,l+1} {\hat b}^{\dagger}_{l} {\hat a}_{l+1} + {\rm h.c.}\right]\,\label{eq6}
\end{eqnarray}
with the emergent density-dependent ``gauge fields'' ${\hat J}^{a/b}_{l,l+1} = \cos[2{\tilde K}_{\Omega} (\hat{n}^{a/b}_{l}-\hat{n}^{a/b}_{l+1})] \exp[i\omega t (\hat{n}^{a/b}_{l}-\hat{n}^{a/b}_{l+1})]$ and ${\hat Q}^{ab/ba}_{l,l+1} = -i\sin[2{\tilde K}_{\Omega} (\hat{n}^{a/b}_{l}-\hat{n}^{b/a}_{l+1})] \exp[i\omega t (\hat{n}^{a/b}_{l}-\hat{n}^{b/a}_{l+1})]$ as well as the renormalized inter-species onsite repulsion ${\hat H}_{\bar U} = {\bar U} \sum_{l} {\hat n}^{a}_{l} {\hat n}^{b}_{l}$.
Obviously, microscopic hopping processes are sorted into two groups: the intra-species and inter-species hoppings between two neighboring sites.
In contrast to the previous work~\cite{wang_2020}, we can see the modulated terms ${\hat J}^{a/b}_{l,l+1}$ and ${\hat Q}^{ab/ba}_{l,l+1}$ in Eq.~(\ref{eq6}) are not pure phases. 
The involved cosine/sine function conserves time-reversal symmetry and belongs to the even/odd sector respectively.
We found that the extra symmetry would help hopping processes select the distinct order of the first-kind Bessel function.

Furthermore, the time-periodic function ${\hat H}_{r}(t)$ can be expanded into a Fourier series ${\hat H}_{r} = \sum^{+\infty}_{-\infty} {\hat H}^{(n)}_{r} \exp(i\omega t)$.
The $0$-th order term or the effective Hamiltonian is given by simply averaging over a period, that is,
\begin{eqnarray}
& & {\hat H}_{e} = {\hat H}^{(0)}_{r} = \frac{1}{T} \int^{T}_{0} dt {\hat H}_{r}(t) = \sum_{l} {\hat h}_{l,l+1}+ {\hat H}_{\bar U}.
\end{eqnarray}
In the derivation, we use a relation of ${\cal J}_{-1}(x) = -{\cal J}_{1}(x) = {\cal J}_{1}(-x)$.
For the intra-species/inter-species hoppings governed by ${\hat J}^{a/b}_{l,l+1}$/${\hat Q}^{ab/ba}_{l,l+1}$, the time-average of the cosine/sine term in a period leads to identity/${\cal J}_{1}$ separately.

\section{Perturbation theory}\label{AppB}
To obtain the low-energy effective Hamiltonian $\hat{H}_{e}$ in a subspace ${\cal H}_{0}$, we follow the strong-coupling expansion method suggested by Takahashi~\cite{Takahashi_1977}.
For a system described by a Hamiltonian \(\hat{H} = \hat{H}_{0} + \hat{V}\), the expansion takes the form
\begin{equation}
  \label{eq:heff}
  \begin{aligned}
  \hat{H}_{e} &= \mathcal{P}_{0} \hat{H}_{0} \mathcal{P}_{0} + \hat{H}^{(1)} + \hat{H}^{(2)} + \cdots \\
  \end{aligned}
\end{equation}
where
\begin{eqnarray}
  \begin{aligned}
  \hat{H}^{(1)} = \mathcal{P}_{0} {\hat V} \mathcal{P}_{0}\ ,\  \hat{H}^{(2)} = \mathcal{P}_{0} {\hat V} \frac{\mathcal{P}_{0} - 1}{E_{0} - \hat{H}_{0}} {\hat V} \mathcal{P}_{0}\ ,
  \end{aligned}
\end{eqnarray}
with the projection operator \(\mathcal{P}_{0}\) onto $H_0$ and the zeroth order energy \(E_{0}\).

Firstly, let us consider the case of zero $J_{0}$ but small $J_{1} \ll {\bar U}$.
At integer-$1$ filling, the groundstate for the non-perturbed \(\hat{H}_{0}\equiv\hat{H}_{\bar{U}}\) is \(2^L\)-fold degenerate with \(E_0 = 0\).
Similar to the Hubbard model, all the odd-order perturbation terms vanish~\cite{Takahashi_1977}.
So we obtain the effective Hamiltonian (\ref{eq:2rd}) up to the second order using the following identities
\begin{equation}
  \begin{aligned}
    \mathcal{P}_{0} \hat{a}_{l}^{\dagger}\hat{b}_{l}\mathcal{P}_{0} &= \hat{S}_{l}^{+}\ , \quad \mathcal{P}_{0}\hat{a}_{l}^{\dagger}\hat{a}_{l}\mathcal{P}_{0} = \frac{1}{2} + \hat{S}_{l}^{z}\ , \\
    \mathcal{P}_{0} \hat{a}_{l}\hat{b}_{l}^{\dagger}\mathcal{P}_{0} &= \hat{S}_{l}^{-}\ , \quad \mathcal{P}_{0}\hat{b}_{l}^{\dagger}\hat{b}_{l}\mathcal{P}_{0} = \frac{1}{2} - \hat{S}_{l}^{z}\ .
  \end{aligned}
\end{equation}

\begin{figure}[b]
\includegraphics[width=0.95\columnwidth]{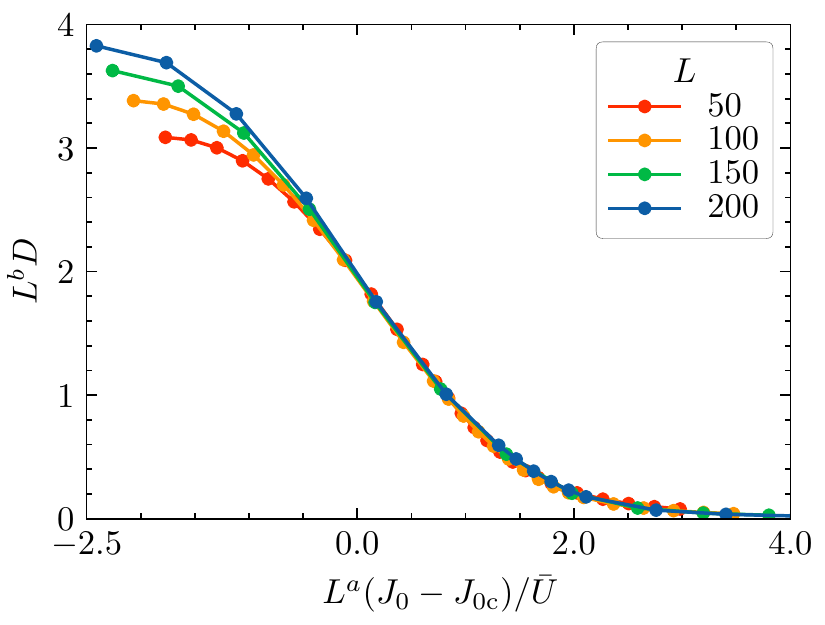}
\caption{
A finite-size analysis of the dimerization strength $\cal D$ at $J_{1} / {\bar U}=1.5$ following the standard scaling theory with polynomial functions.
For $L=50$ (red), $100$ (orange), $150$ (green) and $200$ (blue), the best fitting to the raw data suggests a critical point $J^{c}_{0}=1.7724$ with exponents $a = 0.2213$ and $b=0.1988$.
}\label{fig5}
\end{figure}

As $J_{0}=0$ and $J_{1} \gg {\bar U}$, we decouple odd and even bonds for generating the dimerized manifold naturally.
The Hamiltonian \(\hat{H}_{0} \equiv -J_{1} \sum_{\mathrm{even}\, l} [\hat{H}_{l,l+1}^{ab} + \hat{H}_{l,l+1}^{ba}] + \hat{H}_{\bar{U}}\) for all even bonds is exactly solvable.
The corresponding groundstate wave function $|\psi_{0} \rangle = \otimes_{\mathrm{even}\, l} |\phi_{l,l+1}\rangle$ is a product of local dimers $|\phi_{l,l+1}\rangle = B[ 4 J_{1} (|ab\rangle_{l} |0\rangle_{l+1} + |0\rangle_{l} | ab \rangle_{l+1}) + (\bar{U}+A) (| a \rangle_{l}| a \rangle_{l+1} + | b \rangle_{l}| b \rangle_{l+1}) ]$ with \(A = \sqrt{16 J_{1}^{2} + \bar{U}^{2}}\) and \(B = 1/ (2\sqrt{A^{2} + \bar{U}A})\).
The resulting energy is \(E_{0}= L \left( \bar{U} - A \right) / 2\).
We treat the Hamiltonian $\hat{V} \equiv -J_{1} \sum_{\mathrm{odd}\, l} [\hat{H}_{l,l+1}^{ab} + \hat{H}_{l,l+1}^{ba}]$ for all odd bonds as the perturbation.
It is easy to know that the projection of ${\hat V}$ onto the subspace \(\mathcal{H}_{0}\) in ${\hat H}^{(1)}$ is always null.
The leading-order term comes from the second-order perturbation.
At last, the second-order perturbation term ${\hat H}^{(2)}$ is simplified by \(\langle \psi | 1/(E_{0}-\hat{H}_{0}) | \psi \rangle \approx 1/[E_{0} - \langle \psi|\hat{H}_{0}| \psi \rangle]\) and the energy modification is \(E_{0}^{(2)} \approx \langle \psi_{0}| {\hat V}^{2} | \psi_{0} \rangle / E_{0} = L J_{1}^{2} [64 J_{1}^{4} + {({\bar U} + A)}^{4}]/[2(A^{2}+{\bar U}A)^{2}({\bar U}-A)]\).

\section{Finite-size analysis}\label{AppC}
In this section, we check the scaling behavior of dimerization strength $\cal D$ in the vicinity of the SF-DM transition and attempt to find out whether it follows the ordinary finite-size scaling theory with polynomial functions.

For example, we fix $J_{1}/{\bar U} = 1.5$ and complete the standard finite-size analysis in Fig.~\ref{fig5}. We can see that all curves are nearly overlapping on the SF side while their difference remains finite on the DM side, so called the {\it imbalanced data collapse}, which was considered as a signal of the BKT transition in many previous works~\cite{Pai_2005, Ceccarelli_2013, Dalmonte_2015, Sun_2015, wang_2021}.
In addition, near the suggested critical point $J^{c}_{0}=1.7724$, the system still behaves like the SF phase, which has vanishing $\Delta_{c}$, $\Delta_{n}$ and $\cal D$ as well as a convergent $\rho_s$ independent of the system size $L$ in Fig.~\ref{fig4}.
Further confirmed by the unusual critical exponents $a = 0.2213$ and $b=0.1988$ in Fig.~\ref{fig5}, we suppose that the SF-DM transition is of the BKT type.

\bibliography{tcdb}

\end{document}